\documentclass[twocolumn, journal]{IEEEtran}

\IEEEoverridecommandlockouts

\usepackage{color}
\usepackage{graphicx}
\usepackage{amsmath}
\usepackage{amssymb}
\usepackage{algorithm}
\usepackage{algorithmic}
\usepackage{amsmath}
\usepackage{multirow}
\usepackage{booktabs}
\usepackage{array}
\usepackage{amsthm}
\usepackage{stfloats}
\usepackage{caption}
\usepackage{bm}
\usepackage{tabulary}

\newcommand{\be}{\begin{equation}}
\newcommand{\ee}{\end{equation}}
\newcommand{\bea}{\begin{eqnarray}}
\newcommand{\eea}{\end{eqnarray}}
\newcommand{\ba}{\begin{array}}
\newcommand{\ea}{\end{array}}

\makeatletter

\newcommand{\Rmnum}[1]{\expandafter\@slowromancap\romannumeral #1@}
\makeatother

\title{Non-Cooperative Resource Management for Intelligent Reflecting Surface Aided Networks
\thanks{W. Cai and M. Li are with the School of Information and Communication Engineering, Dalian University of Technology, Dalian 116024, China (e-mail: wenhaocai@mail.dlut.edu.cn; mli@dlut.edu.cn).}
\thanks{Q. Liu is with the School of Computer Science and Technology, Dalian University of Technology, Dalian 116024, China (e-mail: qianliu@dlut.edu.cn).}}
\author{Wenhao Cai,
        Ming Li,~\IEEEmembership{Senior Member,~IEEE,}
        and Qian Liu,~\IEEEmembership{Member,~IEEE}\vspace{-0.1 cm}}

\begin{document}
\maketitle
\begin{abstract}
Intelligent reflecting surface (IRS) has emerged as a promising and revolutionizing technology for future wireless networks. Most existing IRS studies focus on simple cooperative systems which usually have a single frequency band. In realistic non-cooperative multi-band networks, however, the existing IRS designs may be not applicable or have severe performance degradation. Thus, in the complex network environment, it is more rational to consider IRSs as public resources to be dynamically allocated to appropriate users. In this paper, we first introduce the auction theory to tackle the resource management problem for a multi-IRS-assisted non-cooperative network. An efficient auction algorithm framework is introduced to sub-optimally solve this non-convex problem. Simulation result illustrates that the significant performance improvement can be achieved by applying the auction algorithm in the complex multi-IRS-assisted non-cooperative network.
\end{abstract}


\maketitle
\section{Introduction}
The innovative concept of intelligent reflecting surface (IRS) is a promising and revolutionizing technology \cite{Online 1}, which is able to intelligently change propagation environment.
Existing researches have illustrated the advantage of an IRS-assisted communication system, which has one base station (BS) serving multiple users with the aid of one IRS \cite{IRS_spectral}, \cite{Two-stage}.
Nevertheless, realistic wireless networks are usually more complex, which consist of multi-BS, multi-user, and multi-IRS.
In \cite{IRS_multi-cell2}, \cite{IRS_multi-cell4}, the authors considered the joint beamforming, phase-shift, resource allocation, and user association optimization in the IRS-assisted multi-cell networks.
In \cite{IRS-cell-free}, the authors jointly designed beamforming and phase-shift in the IRS-assisted cell-free communication system.
However, IRS control and computation costs brought by the joint optimization methods are severe and unaffordable in practice, which hinders the deployment of IRSs in practical communication networks.
Similar to the IRS-aided mobile edge computing (MEC) \cite{MEC}, a promising approach is to first realize IRS resource management (i.e. allocate IRSs), then jointly design the transmit beamforming and phase-shifts at edge computing node, which can significantly reduce the overhead.
Therefore, IRS resource management is significantly important for multi-IRS-assisted wireless networks \cite{IRS_game1}.

Recently, some works have investigated the IRS resource allocation problem \cite{IRS_resource}, \cite{IRS_game2}.
In particular, the authors in \cite{IRS_game1} proposed a \textit{Stackelberg} game in the scenario where BSs and IRSs belong to different service providers to jointly optimize the IRS resource price, transmit power, and IRS passive beamforming.
The authors in \cite{IRS_resource} proposed the parallel alternating direction method of multipliers algorithm (PADMM) to allocate the transmit power and IRS resources to users.
In \cite{IRS_game2}, the authors considered the IRS-assisted wireless network with multiple operators and multiple mobile users. The evolutionary game was proposed to dynamically design the service provider (SP) and service selection strategies for users.

\begin{figure}[t]
\centering
  \includegraphics[height = 1.8 in]{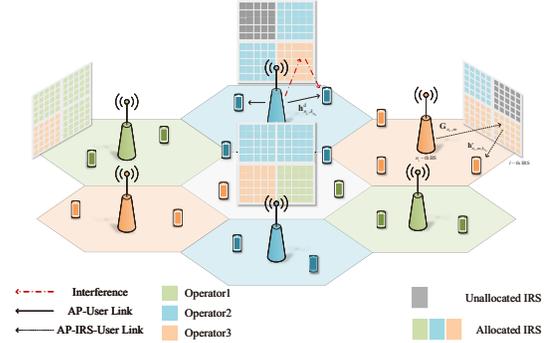}\\
  \vspace{0.1 cm}
  \caption{Multi-IRS-assisted non-cooperative multi-operator communication network.}\label{fig:System model}
  \vspace{-0.6 cm}
\end{figure}

However, all the works mentioned above focus on the scenarios with a cooperative network with a single frequency band.
In practical non-cooperative communication networks using multiple frequency bands, the conventional optimization algorithms are not applicable. This is because the channel state information (CSI) cannot be fully acquired at the IRS controller in non-cooperative networks, which is necessary for the conventional optimization algorithm.
In addition, when we consider the frequency-selective characteristic of the IRS \cite{Practical1}, \cite{Practical2}, the IRS resource allocation problem is more complicated and difficult.
Since an IRS can only provide the desired phase-shift to signals within a certain frequency band, IRSs should be considered as public resources that can be allocated to different operators using different frequency bands.
Thus, the inappropriate allocation of IRS resources will lead to significant spectral/energy efficiency performance degradation.
Moreover, the high computational complexity required for complex calculations and the privacy/security issues associated with data interactions are unacceptable for non-cooperative IRS-aided networks.
Therefore, an efficient algorithm that can design IRS resource allocation in non-cooperative networks urgently needs to be developed, which motivates this work.

%

Inspired by dynamic spectrum auctions \cite{auction}, in this work, for the first time we utilize the auction algorithm to tackle the resource management problem for the multi-IRS-assisted non-cooperative communication network.
We aim to optimize the IRS resource allocation to maximize the sum-rate gain for all IRSs with the auction algorithm, which can maximize IRS utilization by introducing market competition into the non-cooperative communication network. Specifically, in the non-cooperative communication network, the IRS service provider sets the auction rules. Then, different operators need to adequately bid to maximize their own profits. After a few rounds of bidding, the IRS resource allocation will reach a steady-state.
An efficient auction algorithm framework is introduced to sub-optimally solve this non-convex problem.
Simulation result illustrates that significant performance improvement can be achieved by applying the auction algorithm in the non-cooperative multi-operator communication network.

\section{System Model and Problem Formulation}
We consider a multi-IRS-assisted non-cooperative multi-operator communication network as shown in Fig. \ref{fig:System model}, where $K$ single-antenna users are served by $N$ multi-antennas BSs belonging to $S$ tele-communication operators.
In this multi-operator network, different operators deploy their BSs and provide communication services to their users using different frequency bands. Specifically, each operator has a multi-BS multi-user system, which uses different frequency bands to avoid inter-cell interference.
However, as a device that can change the wireless environment, IRS may affect signals of all BSs in the network.
Since IRS has been widely considered as a frequency-selective device \cite{Practical1}, \cite{Practical2}, each IRS can only provide the desired phase-shift for a certain operator, while it provides uncontrolled phase-shifts for other operators.
It is because the frequency bands used by different operators are significantly different, while the IRS can only effectively adjust the phase-shifts for the signals within a certain frequency range.
For example, the frequency bands of the 4G LTE network are 1880-1900 MHz, 2300-2320 MHz, and 2635-2655 MHz for China Mobile, China Unicom, and China Telecom, respectively.
Thus, the IRS cannot provide effective phase-shifts to the signals of all operators.
Meanwhile, for different BSs, which belong to the same operator and work on similar frequencies, the frequency-selective characteristics are not obvious. Thus, it can be assumed that the IRS can provide the same phase-shift for the BSs belonging to the same operator.

Let $\mathcal{S} \triangleq \{1, \ldots, S\}$ denote the set of operators,
$\mathcal{N}_s \triangleq \{1_s, \ldots, N_s\}$ denote the set of the BSs belonging to the $s$-th operator,
$\mathcal{K}_{n_s} \triangleq \{1_{n_s}, \ldots, K_{n_s}\}$ denote the set of the users served by the $n_s$-th BS,
and $\mathcal{L} \triangleq \{1, \ldots, L\}$ denote the set of IRSs.
Specifically, for each operator, there are $N_s$ multi-antennas BSs serving $K_{n_s}$ single-antenna users simultaneously, where $L$ IRSs are deployed to assist the downlink communication for the multi-operator network.
In addition, the IRS is controlled by a local IRS controller and only the first-order reflection is considered due to significant path loss.

Denote $\mathbf{x}_{n_s} \triangleq [x_{s,1}, \ldots, x_{s,K_{n_s}}]^T \in \mathbb{C}^{K_{n_s}}$ as the transmitted symbols for the $K_{n_s}$ users served by the $n_s$-th BS with $\mathbb{E}\{\mathbf{x}_{n_s} \mathbf{x}_{n_s}^H\} = \mathbf{I}_{K_{n_s}}$, and $\mathbf{W}_{n_s} \triangleq [\mathbf{w}_{n_s,1}, \ldots, \mathbf{w}_{n_s,K_{n_s}}] \in \mathbb{C}^{N_\text{t}\times K_{n_s}}$ as the precoder matrix of the $n_s$-th BS, where $N_\text{t}$ is the number of transmit antennas.
Since BSs operate at different frequencies, the interference can be easily eliminated through receiving filters at the users. Thus, the received baseband signal at the $k_{n_s}$-th user served by the $n_s$-th BS can be expressed as
\begin{equation}
\begin{aligned}
    y_{n_s,k_{n_s}} \triangleq& \Big[ \sum_{l=1}^L(\mathbf{h}^\text{r}_{l,n_s,k_{n_s}})^H\bm{\Theta}_{l,s}
    \mathbf{G}_{l,n_s}+(\mathbf{h}^\text{d}_{n_s,k_{n_s}})^H \Big]\\
    & \times \sum_{j\in\mathcal{K}_{n_s}}\mathbf{w}_{n_s,j}x_{s,k_{n_s}} + z_{s,k_{n_s}},
    \ \forall k_{n_s}, \forall n_s,
    \label{eq:y}
\end{aligned}
\end{equation}
where $z_{s,k_s} \sim \mathcal{C}\mathcal{N}(0,\sigma^2)$ denotes the additive white Gaussian noise (AWGN) at the $k_{n_s}$-th user.
$\bm{\Theta}_{l,s} \triangleq \text{diag}\{\bm{\theta}_{l,s}\}$ and $\bm{\theta}_{l,s} \triangleq [\theta_{l,s,1}, \ldots, \theta_{l,s,M_{l}}]^T$ are the IRS reflection matrices and vectors, respectively, for the signals with specific frequency of the $s$-th operator.
$\mathbf{h}^\text{r}_{l,n_s,k} \in \mathbb{C}^{M_{l}}$, $\mathbf{G}_{l,n_s} \in \mathbb{C}^{M_{l} \times N_\text{t}}$, and $\mathbf{h}^\text{d}_{n_s,k} \in \mathbb{C}^{N_\text{t}}$ represent the baseband equivalent channels from the $l$-th IRS to the $k_{n_s}$-th user, from the $n_s$-th BS to the $l$-th IRS, and from the $n_s$-th BS to the $k_{n_s}$-th user, respectively.
The quasi-static flat-fading Rayleigh channel model is adopted for all channels. We assume that the CSIs of served users and IRSs are perfectly known at the BSs by the existing efficient channel estimation methods \cite{Channel}.

When the frequency-selective characteristic is considered, IRS can effectively adjust phase-shifts for the signals with a certain frequency, while it provides fixed phase-shifts for others \cite{Practical1}, \cite{Practical2}.
Thus, we introduce $a_{l,s} \in \{0,1\}, \forall l, s,$ to indicate if the $l$-th IRS can offer desired phase-shifts for the BSs belonging to the $s$-th operator.
When $a_{l,s} = 0$, $\bm{\Theta}_{l,s}$ is fixed as $\mathbf{I}_{M_l}$ because the phase-shift is uncontrolled \cite{Practical2}. When $a_{l,s} = 1$, $\bm{\Theta}_{l,s}$ is a designable diagonal matrix with a constant module.
Moreover, $\mathbf{a}_l \triangleq [a_{l,1}, \ldots, a_{l,S}]^T, \forall l,$ denotes the $l$-th IRS resource allocation vector, and $||\mathbf{a}_l||_1 = 1$ indicates that each IRS can be allocated to only one operator.
Following this analysis, we establish the set of the tunable IRSs $\mathcal{I}_s \triangleq \{l|a_{l,s} = 1\}$.
Then, with the received signal in (\ref{eq:y}) and under the set of tunable IRSs $\mathcal{I}_s$, the SINR of the $k_{n_s}$-th user served by the $n_s$-th BS is given by
\begin{equation}
\gamma_{k_{n_s}}(\mathcal{I}_s) = \frac{
\big|{\mathbf{h}}_{k_{n_s}}(\mathcal{I}_s)^H\mathbf{w}_{n_s,k_{n_s}}\big|^2}
{\sum^{j \in \mathcal{K}_{n_s}}_{j \neq k_{n_s}}
\big|{\mathbf{h}}_{k_{n_s}}(\mathcal{I}_s)^H\mathbf{w}_{n_s,j}\big|^2+\sigma_{k_{n_s}}^2},
\end{equation}
where
\begin{equation}
\begin{aligned}
\mathbf{h}^H_{k_{n_s}}(\mathcal{I}_s) &\triangleq
   \sum_{l \in \mathcal{I}_s}(\mathbf{h}^\text{r}_{l,n_s,k_{n_s}})^H\bm{\Theta}_{l,s}\mathbf{G}_{l,n_s}\\
+& \sum_{l \in \mathcal{L}\backslash\mathcal{I}_s}(\mathbf{h}^\text{r}_{l,n_s,k_{n_s}})^H\mathbf{I}_{M_l} \mathbf{G}_{l,n_s}
+ (\mathbf{h}^\text{d}_{n_s,k_{n_s}})^H
\end{aligned}
\end{equation}
is defined to represent the combined channel for the $k_{n_s}$-th user with both the tunable IRSs in the set $\mathcal{I}_s$ and other fixed IRSs for the $s$-th operator.
Furthermore, if there is no IRS serving the $s$-th operator, i.e., $\mathcal{I}_s = \varnothing$, the combined channel degenerates into
\begin{equation}
\mathbf{h}^H_{k_{n_s}}(\varnothing) \triangleq\sum_{l \in \mathcal{L}}
(\mathbf{h}^\text{r}_{l,n_s,k_{n_s}})^H\mathbf{I}_{M_l} \mathbf{G}_{l,n_s}
+ (\mathbf{h}^\text{d}_{n_s,k_{n_s}})^H,
\end{equation}
and the resulting $\gamma_{k_{n_s}}(\varnothing)$ denotes the SINR of the $k_{n_s}$-th user without the assistance by any tunable IRSs.

Our goal is to jointly optimize the transmit beamformers $\mathbf{W}_{n_s}, \forall {n_s} \in \mathcal{N}_s,$ the IRS phase-shifts $\bm{\Theta}_{l,s}, \forall l \in \mathcal{I}_s, \forall s \in \mathcal{S},$ and the IRS resource allocation matrix $\mathbf{A} \triangleq [\mathbf{a}_1, \ldots, \mathbf{a}_L]$ to maximize the sum-rate gain brought by $L$ IRSs, subject to the transmit power budget of each BS and the IRS service restrictions. This problem can be expressed as
\begin{subequations}
\label{eq:problem}
\begin{align}
   \label{eq:problem a}
   \max\limits_{\mathbf{A},\bm{\Theta}_{l,s},\mathbf{W}_{n_s}}~&
   \sum_{s=1}^S\sum_{n_s=1}^{N_s}\sum_{k_{n_s}=1}^{K_{n_s}}\Big\{
   \log[1+\gamma_{k_{n_s}}(\mathcal{I}_s)]\nonumber\\
   &~~~~~~~~~~~~~~~~~~~~-\log[1+\gamma_{k_{n_s}}(\varnothing)]\Big\}\\
   \label{eq:problem b}
   \textrm{s.t.}~~&\sum_{k_{n_s}}\left\|\mathbf{w}_{n_s,k_{n_s}}\right\|^2 \leq P_{n_s},\;\;\forall n_s,\\
   \label{eq:problem c}
   &\left\| \mathbf{a}_l \right\|_1 = 1, \;\;a_{l,m} \in \{0,1\}, \;\;\forall l,\; m,
\end{align}
\end{subequations}
where $P_{n_s} > 0$ denotes the transmit power budget of the $n_s$-th BS.
The non-convex NP-hard problem (\ref{eq:problem}) is very difficult to solve due to the following reasons.
Firstly, complete CSIs are unavailable for all non-cooperative operators to implement joint design of beamforming, phase-shifts, and IRS resource allocation. This is because each operator can only obtain the CSIs of served users and allocated IRSs.
Secondly, the complicated objective function (\ref{eq:problem a}) and the discrete constraint (\ref{eq:problem c}) make this multi-variable problem difficult to solve using the existing convex optimization methods.

Inspired by the dynamic spectrum auctions \cite{auction}, we utilize the auction theory to tackle the IRS allocation problem, by which IRS resources will be most efficiently allocated by the market rather than the inefficient central controller.
In addition, an IRS can only provide phase-shift to the signals from a certain operator because of its frequency-selective characteristic (\ref{eq:problem c}). This fact further motivates the use of the auction method since each IRS can be allocated to only one operator.
More importantly, with the aid of auction theory, the operators only needs to interact the price information rather than all CSIs, which is indispensable in such a non-cooperative communication network design.
After obtaining the optimal IRS resource allocation, the transmit beamformers and the IRS phase-shifts can be easily designed by the existing local algorithms \cite{IRS_spectral}-\cite{IRS-cell-free}, which will be omitted due to space limitation.

\section{IRS Allocation with Auction Theory}
In this section, we propose to use the auction theory to efficiently design the IRS resource allocation.
To introduce auction theory to the non-cooperative network, we first define the utility function.
Then, in order to find a suitable auction rule to solve the original problem, we compare different auction mechanisms.
Finally, we proposed two appropriate auction algorithms for the IRS resource allocation and introduced a framework to design the joint optimizing problem for this non-cooperative network.

\subsection{Utility Functions}
For a certain operator, the performance gain brought by the allocated IRSs (i.e., $\forall l \in \mathcal{I}_s$) can be expressed as
\begin{equation}
  g(\mathcal{I}_s) \triangleq \hspace{-0.1cm} \sum_{n_s=1}^{N_s} \hspace{-0.05cm} \sum_{k_{n_s}=1}^{K_{n_s}} \hspace{-0.1cm}
  \Big\{\hspace{-0.05cm}\log[1+\gamma_{k_{n_s}}(\mathcal{I}_s)]-\log[1+\gamma_{k_{n_s}}(\varnothing)]\hspace{-0.05cm}\Big\}.
\end{equation}
Although the operator needs to evaluate the value of each IRS before auction, it is difficult to realize the real-time update of the value during the auction. Instead, the $s$-th operator utilizes the sum-rate gain between the scenario with only the $l$-th IRS (i.e., $a_{s,l} = 1$) and the scenario without any IRSs (i.e., $a_{s,l} = 0, \forall l$) as the value of the $l$-th IRS for $s$-th operator.
Thus, in order to solve this problem more efficiently, we assume that the $s$-th operator can evaluate the value of the $l$-th IRS by
\begin{equation}\label{nu}
\nu_{l,s} \triangleq \hspace{-0.1cm} \sum_{n_s=1}^{N_s}\sum_{k_{n_s}=1}^{K_{n_s}} \hspace{-0.1cm}
\Big\{\hspace{-0.05cm}\log[1+\gamma_{k_{n_s}}(\{l\})]-\log[1+\gamma_{k_{n_s}}(\varnothing)]\Big\}.
\end{equation}

\subsection{Auction Rules}
After defining the utility functions, it is necessary to find a suitable auction rule to maximize auction benefits, ensure the fairness, and avoid the winner's curse.
Thus, we focus on various available auction algorithms, which mainly includes the following mechanisms.
\begin{enumerate}
  \item Bidding direction:
        The British-style auction is known as an advance auction, which follows the principle of highest bidder wins.
        On the contrary, the bids are descending in the Dutch auction. In the IRS auction, the Dutch auction is not appropriate because of its low efficiency, high overhead, and the difficulty of compatibility with complementary IRS resources.
  \item Bidding information:
        The public auction intensifies the competition among participants, which allows the market to function. This auction rule requires a few data exchange among IRS provider and different operators, which is affordable.
        On the contrary, in the closed auction, all participants know only their own information. This auction rule requires the least information exchange between operators and IRS providers.
  \item Bidding order:
        For simultaneous auction, bidders simultaneously bid for the licenses of all IRSs, which has high efficiency.
        For successive auction, multiple IRS licenses are bidden one after another. Thus, each round of the auction is easy and consumes few resources.
        Particularly, simultaneous auctions only make sense in the context of public auctions, as bidders must adjust their bids to the reactions of other participants.
\end{enumerate}
In this work, two auction algorithms are developed with the aim of solving the original problem (\ref{eq:problem}).
The detailed auction algorithms are given in the following subsections.

\subsection{Successive Advance Auction}
In the successive advance auction, all IRSs are auctioned one by one.
The $s$-th operator (i.e., bidder) estimates the profit of the $l$-th IRS by
\begin{equation}\label{rho}
    \rho_{l,s} \triangleq \max\big\{\nu_{l,s} - \epsilon_{l}, 0\big\}, \forall l, \forall s,
\end{equation}
where $\epsilon_{l} = \max\{\beta_{l,s}, \forall s\}$ denotes the current price of the $l$-th IRS, which is determined by the highest bids among $S$ operators after each auction round.
Then, if there is any un-allocated IRS (i.e., $a_{s,l} = 0, \forall s$) with a positive profit, the operator always chooses the IRS with the maximum profit, and it offers an appropriate bid as
\begin{equation}\label{beta}
    \beta_{l,s} \triangleq \epsilon_{l} + \kappa(\rho_{l,s} - \epsilon_{l}),
\end{equation}
where $\kappa \in (0,1)$ denotes the bidding coefficient.
The coefficient $\kappa$ determines the convergence speed and performance of the auction algorithm, and is selected as $\kappa = \frac{1}{S}$ in this work.
The auction ends when all IRSs have no profit to all operators, i.e.,
$\rho_{l,s} = 0, \forall l, \forall s.$
Finally, the $l$-th IRS can be allocated by
\begin{equation}\label{A}
a_{l,s}=\left\{
\begin{aligned}
& 1,  \epsilon_{l} = \beta_{l,s},\\
& 0,  \text{otherwise}.
\end{aligned}
\right.
\end{equation}
The successive advance auction algorithm is summarized in Algorithm 1.

\begin{algorithm}[t]\footnotesize
\caption{The Successive Advance Auction Algorithm.}
    \begin{algorithmic}[1]
    \begin{small}
    \REQUIRE $\mathbf{h}^\text{r}_{l,n_s,k_{n_s}}, \mathbf{G}_{l,n_s}, \mathbf{h}^\text{d}_{n_s,k_{n_s}}, \sigma_{k_{n_s}}, \bm{\Theta}_{l,s},\mathbf{W}_{n_s}, \nu_{l,s}$.
    \ENSURE $\mathbf{A}$.
    \WHILE{$\rho_{l,s} = 1, \exists \ l, s,$}
    \FOR{$s=1$ to $S$}
        \FOR{$l=1$ to $L$}
            \STATE{Calculate $\rho_{l,s}$ by (\ref{rho}).}
        \ENDFOR
        \STATE{Calculate $\beta_{l,s}$ by (\ref{beta}).}
    \ENDFOR
    \STATE{Calculate $\epsilon_{l} = \max\{\beta_{l,s}, \forall s\}$ and $a_{l,s}$ by (\ref{A}).}
    \ENDWHILE
    \STATE {Return $\mathbf{A}^\star$.}
    \end{small}
    \end{algorithmic}
\end{algorithm}

\begin{algorithm}[t]\footnotesize
\caption{The Simultaneous Multi-round Auction Algorithm.}
    \begin{algorithmic}[1]
    \begin{small}
    \REQUIRE $\mathbf{h}^\text{r}_{l,n_s,k_{n_s}}, \mathbf{G}_{l,n_s}, \mathbf{h}^\text{d}_{n_s,k_{n_s}}, \sigma_{k_{n_s}}, \bm{\Theta}_{l,s},\mathbf{W}_{n_s}, \nu_{l,s}$.
    \ENSURE $\mathbf{A}$.
    \WHILE{no convergence of $\mathbf{A}$}
    \FOR{$s=1$ to $S$}
        \FOR{$l=1$ to $L$}
            \STATE{Calculate $\rho_{l,s}$ by (\ref{rho}).}
        \ENDFOR
        \STATE{Calculate $\overline{\rho}_{s}$ by (\ref{orho}).}
        \STATE{Calculate $\beta_{l,s}$ by (\ref{beta2}).}
    \ENDFOR
    \STATE{Calculate $\epsilon_{l} = \max\{\beta_{l,s}, \forall s\}$ and $a_{l,s}$ by (\ref{A}).}
    \ENDWHILE
    \STATE {Return $\mathbf{A}^\star$.}
    \end{small}
    \end{algorithmic}
\end{algorithm}

\subsection{Simultaneous Multi-round auction}
In the simultaneous multi-round auction, operators simultaneously bid multiple IRSs. First, the $s$-th operator calculates the profit of $l$-th IRS by (\ref{rho}), and the average profit can be calculated by
\begin{equation}\label{orho}
    \overline{\rho}_{s} \triangleq \frac{\sum_{l=1}^L\rho_{l,s}}{L}, \forall s.
\end{equation}
Then, for any appropriate IRS which has a higher profit $\rho_{l,s}$ than $\overline{\rho}_{s}$, the $s$-th operator offers a bid as
\begin{equation}\label{beta2}
    \beta_{l,s} \triangleq \rho_{l,s} - \overline{\rho}_{s}.
\end{equation}
Similarly, the $l$-th IRS can be allocated by (\ref{A}). Then, the auction ends when the IRS resource allocation matrix $\mathbf{A}$ reaches convergence. The simultaneous multi-round auction algorithm is summarized in Algorithm 2.

\subsection{Summary}
In this work, we propose two auction algorithms to design IRS resource allocation in the non-cooperative multi-operator communication network.
The simultaneous multiple round auction converges more quickly, while successive advance auction has better performance with an appropriate $\kappa$.
The comparison of different proposed algorithms is summarized in Table I.
In practice, operators first initialize the transmit beamformers and phase-shifts for all IRSs by conventional local algorithms \cite{IRS_spectral}-\cite{IRS-cell-free}.
Then, operators calculate the value of IRSs and the IRS provider design IRS resource allocation by the proposed auction algorithms.
Finally, the operators design the transmit beamformers and phase-shifts for the allocated IRSs by conventional local algorithms again.
The proposed auction-based algorithm framework is summarized in Algorithm 3.

\begin{algorithm}[t]\footnotesize
\caption{The Auction Based Algorithm Framework.}
\label{alg:SH}
    \begin{algorithmic}[1]
    \begin{small}
    \REQUIRE $\mathbf{h}^\text{r}_{l,n_s,k_{n_s}}, \mathbf{G}_{l,n_s}, \mathbf{h}^\text{d}_{n_s,k_{n_s}}, \sigma_{k_{n_s}}, \forall s, n_s, k_{n_s}$.
    \ENSURE $\mathbf{A}, \bm{\Theta}_{l,s},\mathbf{W}_{n_s}, \forall l, s, n_s,$.
    \FOR{$s=1$ to $S$}
        \STATE{Initialize $\bm{\Theta}_{l,s},\mathbf{W}_{n_s}, \forall l, s, n_s,$ by local algorithm.}
    \ENDFOR
    \FOR{$l=1$ to $L$}
        \FOR{$s=1$ to $S$}
            \STATE {Calculate $\nu_{l,s}$ by (\ref{nu}).}
        \ENDFOR
    \ENDFOR
    \STATE {Calculate $\mathbf{A}^\star$ by auction algorithm.}
    \FOR{$s=1$ to $S$}
        \STATE{Calculate $\bm{\Theta}^\star_{l,s},\mathbf{W}^\star_{n_s}, \forall l, s, n_s,$ by local algorithm.}
    \ENDFOR
    \STATE {Return $\mathbf{A}^\star, \bm{\Theta}^\star_{l,s},\mathbf{W}^\star_{n_s}, \forall l, s, n_s$.}
    \end{small}
    \end{algorithmic}\vspace{-0.0 cm}
\end{algorithm}

\begin{table}[t]
	\centering
	\caption{The comparison of proposed algorithms.}
	\label{table1}
	\begin{tabular}{c|ccc}
		\hline  
		& & & \\[-6pt]
		                    &Successive&Simultaneous&IRS \\
        Different algorithm &advance   &multi-round &exhaustive \\
                            &auction   &auction     &search \\
		\hline
		& & & \\[-6pt]
		Bidding direction     &\textit{British-style} &\textit{British-style} &- \\
		\hline
		& & & \\[-6pt]
		Bidding information   &private  &public  &-\\
		\hline
		& & & \\[-6pt]
		Bidding order         &successive &simultaneous   &- \\
		\hline
		& & & \\[-6pt]
		Signaling overhead  &Low        &Low        &High \\
		\hline
		& & & \\[-6pt]
		Distributed method    &\checkmark &\checkmark &$\times$\\
		\hline
		& & & \\[-6pt]
		{Complexity}
        &{$\mathcal{O}\{S(L+1)\}$}
        &{$\mathcal{O}\{S(L+2)\}$}
        &{$\mathcal{O}\{S^L\}$}\\
		\hline
	\end{tabular}
\end{table}

\section{Simulation Result}
In this section, simulation results are presented to demonstrate the significance of using the auction theory in the non-cooperative multi-operator communication network.
We assume that the non-cooperative multi-operator network consists of $S=3$ operators.
Each operator has $N_s = 2, \forall s,$ BSs equipped with $N_\text{t} = 8$ antennas.
Each BS serves $K = K_{n_s} = 3, \forall {n_s},$ single-antenna users, where $L = 6$ IRSs composed of $M = M_l = 64, \forall l,$ reflecting elements are randomly distributed near the users to assist the downlink communications.
The noise power is set as $\sigma^2 = -70$dBm.
The power budgets of each BS are the same, i.e., $P = P_{n_s}=-5\text{dBW}, \forall {n_s}$.
In addition, the distance-dependent channel path loss is modeled as $\eta(d) = C_0(\frac{d}{d_0})^{-\alpha}$, where $C_0 = -30$dB denotes the signal attenuation at the reference distance $d_0 = 1$m, and $\alpha$ denotes the path loss exponents.
We set the path-loss exponents for the BS-IRS, IRS-user, and BS-user channels as 2.5, 2.8, and 3.5, respectively.

\begin{figure}[t]
\centering
  \includegraphics[height = 2.5 in]{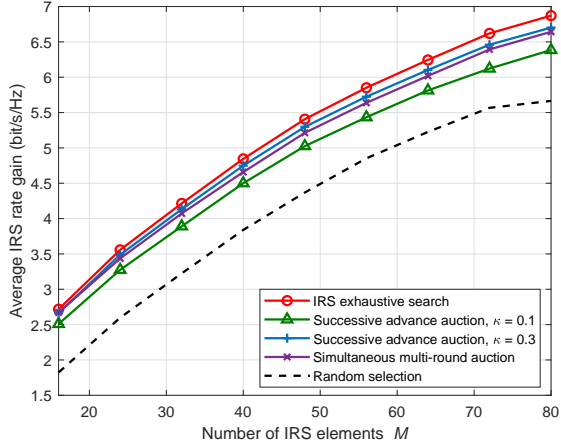}
  \vspace{-0.1cm}
  \caption{The average IRS rate gain versus number of IRS elements $M$ ($L$ = 6).}\label{fig:result}
  \vspace{-0.4cm}
\end{figure}

\begin{figure}[t]
\centering
  \includegraphics[height = 2.5 in]{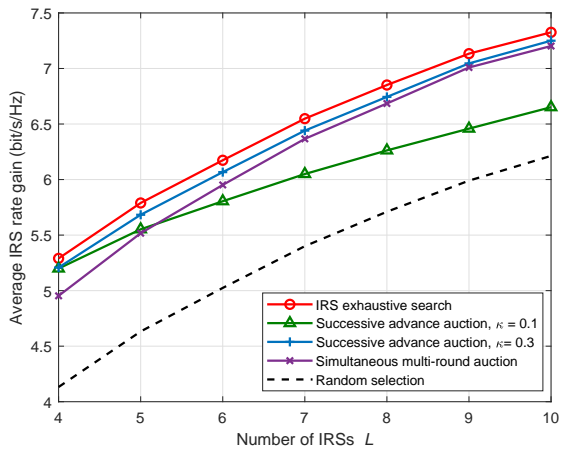}
  \vspace{-0.1cm}
  \caption{The average IRS rate gain versus number of IRSs $L$  ($M = 64$).}\label{fig:result2}
  \vspace{-0.4cm}
\end{figure}

Fig. \ref{fig:result} shows the average IRS rate gain versus the number of IRS elements $M$ by using the proposed auction algorithms for optimizing the non-cooperative IRS-assisted network.
Our algorithms based on the auction theory in this paper are denoted as ``Successive advance auction'' and ``Simultaneous multi-round auction''.
For comparison, we also include
\textit{i)} the scenario that all IRSs are allocated by exhaustive search, which is denoted as ``IRS exhaustive search'' and serves as a performance upper bound;
\textit{ii)} the scheme that each IRS is randomly allocated to a BS, which is denoted as ``Random selection''.
It is obvious that the proposed algorithms achieve IRS rate gain approaching to the ``IRS exhaustive search" and significantly greater than ``Random selection", which validates the advantages of allocating IRSs with auction theory.

Next, the average IRS rate gain versus the number of IRSs $L$ is plotted in Fig. \ref{fig:result2}. A similar conclusion can be drawn as that from Fig. \ref{fig:result}.
In addition, it can be observed that the curves with the legend ``Successive advance auction, $\kappa = 0.3$'', ``Successive advance auction, $\kappa = 0.1$'' have different performances. This is because that the bidding coefficient $\kappa$ determines the convergence speed and performance of the auction algorithm, and our proposed $\kappa = \frac{1}{S}$ is an appropriate selection.

\section{Conclusions}
In this paper, we proposed to utilize the auction theory to solve the non-cooperative resource management problem for the multi-IRS-assisted communication network.
We formulated a problem to jointly optimize the transmit beamformers, IRS phase-shifts, and IRS resource allocation matrix to maximize the sum-rate gain brought by all IRSs.
An efficient auction algorithm framework has been introduced to sub-optimally solve this non-convex problem.
Simulation results illustrated that the significant performance improvement can be achieved by applying the auction algorithm in the non-cooperative multi-operator communication network.


\begin{thebibliography}{99}
\bibitem{Online 1} Q. Wu and R. Zhang, ``Towards smart and reconfigurable environment: Intelligent reflecting surface aided wireless network,'' \textit{IEEE Commun. Mag.}, vol. 58, no. 1, pp. 106-112, Jan. 2020.
\bibitem{IRS_spectral} S. Zhou, W. Xu, K. Wang, M. D. Renzo, and M. Alouini, ``Spectral and energy efficiency of IRS-assisted MISO communication with hardware impairments,'' \textit{IEEE Wireless Commun. Lett.}, vol. 9, no. 9, pp. 1366-1369, Sep. 2020.
\bibitem{Two-stage} Q. Wu and R. Zhang, ``Intelligent reflecting surface enhanced wireless network via joint active and passive beamforming design,'' \textit{IEEE Trans. Wireless Commun.}, vol. 18, no. 11, pp. 5394-5409, Nov. 2019.
\bibitem{IRS_multi-cell2} Y. Jia, C. Ye, and Y. Cui, ``Analysis and optimization of an intelligent reflecting surface-assisted system with interference," \textit{IEEE Trans. Wireless Commun.}, vol. 19, no. 12, pp. 8068-8082, Dec. 2020.
\bibitem{IRS_multi-cell4} C. Pan, H. Ren, K. Wang, W. Xu, M. Elkashlan, A. Nallanathan, and L. Hanzo, ``Multicell MIMO communications relying on intelligent reflecting surfaces," \textit{IEEE Trans. Wireless Commun.}, vol. 19, no. 8, pp. 5218-5233, Aug. 2020.
\bibitem{IRS-cell-free} Z. Zhang and L. Dai, ``A joint precoding framework for wideband reconfigurable intelligent surface-aided cell-free network," \textit{IEEE Trans. Signal Process.}, vol. 69, pp. 4085-4101, Jun. 2021.
\bibitem{MEC} T. Bai, C. Pan, Y. Deng, M. Elkashlan, A. Nallanathan, and L. Hanzo, ``Latency minimization for intelligent reflecting surface aided mobile edge computing," \textit{IEEE J. Sel. Areas Commun.}, vol. 38, no. 11, pp. 2666-2682, Nov. 2020.
\bibitem{IRS_game1} Y. Gao, C. Yong, Z. Xiong, D. Niyato, Y . Xiao, and J. Zhao, ``A stackelberg game approach to resource allocation for intelligent reflecting surface aided communications,''
    in \textit{Proc. IEEE Global Commun. Conf. (GLOBECOM)}, Taipei, Taiwan, Dec. 2020.
\bibitem{IRS_resource}Y. Gao, C. Yong, Z. Xiong, J. Zhao, Y. Xiao, and D. Niyato, ``Reflection resource management for intelligent reflecting surface aided wireless networks," \textit{IEEE Trans. Commun.,} vol. 69, no. 10, pp. 6971-6986, Oct. 2021.
\bibitem{IRS_game2} N. C. Luong, N. T. T. Van, S. Feng, H. T. Nguyen, D. Niyato, and D. I. Kim, ``Dynamic network service selection in IRS-assisted wireless networks: A game theory approach,'' \textit{IEEE Trans. Veh. Tech.}, vol. 70, no. 5, pp. 5160-5165, Apr. 2021.
\bibitem{Practical1} H. Li, W. Cai, Y. Liu, M. Li, Q. Liu, and Q. Wu, ``Intelligent reflecting surface enhanced wideband MIMO-OFDM communications: From practical model to reflection optimization,'' \textit{IEEE Trans. Commun.}, vol. 69, no. 7, pp. 4807-4820, Mar. 2021.
\bibitem{Practical2} W. Cai, R. Liu, Y. Liu, M. Li, and Q. Liu, ``Intelligent reflecting surface assisted multi-cell multi-band wireless networks,'' in \textit{Proc. IEEE Wireless Commun. Network Conf. (WCNC)}, Nanjing, China, Mar. 2021.
\bibitem{auction} F. Benedetto, L. Mastroeni, and G. Quaresima, ``Auction-based theory for dynamic spectrum access: A review," in \textit{Proc. Int. Conf. Telecommun. Signal Process.}, Jul. 2021, pp. 146-151.
\bibitem{Channel} A. L. Swindlehurst, G. Zhou, R. Liu, C. Pan, and M. Li, ``Channel estimation with reconfigurable intelligent surfaces -- A general framework," \textit{Proceedings of the IEEE}, to appear.
\end{thebibliography}
\end{document}